\documentclass[prl,showpacs, twocolumn,groupedaddress]{revtex4}

\usepackage{graphicx}

\begin{document}

\title
{The Classical-Map Hyper-Netted-Chain (CHNC) technique
 for inhomogeneous electron systems. Application to quantum dots.
}

\author{M. W. C. Dharma-wardana}
\email{chandre.dharma-wardana@nrc- cnrc.gc.ca}
\affiliation{
Institute for Microstructural Sciences, National Research Council of Canada,
Ottawa, Canada K1A 0R6}

\date{\today}
\begin{abstract}
The Classical-map Hyper-Netted-Chain (CHNC) technique is a simple method
of calculating quantum pair-distribution functions, spin-dependent energies,
etc., of strongly-interacting {\it uniform} systems. We present CHNC
calculations of charge densities and energies of {\it non-uniform} systems,
viz., quantum dots, and compare with quantum Monte Carlo and density
-functional results.  Results
for up to 210 electrons are reported.

\end{abstract}
\pacs{PACS Numbers: 71.10.Lp,75.70.Ak,73.22-f}
%
\maketitle
%
The Hohenberg-Kohn and Mermin (HKM) theorems~\cite{hohen} of density-functional
theory (DFT) assert that the
one-body density $n(\vec{r})$ of an inhomogeneous system completely
determines its physics. However,
 DFT uses the more laborious Kohn-Sham (K-S)
approach~\cite{kohnsham} due to the lack of an accurate kinetic-energy
 functional~\cite{PerrotH}. 
The Kohn-Sham $n(\vec{r})$ of an electron system is:
\begin{equation}
n(\vec{r})=\sum_\nu |\phi_\nu(\vec{r})|^2f_\nu(\epsilon_\nu/T)
\end{equation}
Here $\phi_\nu$ are K-S eigenstates with ``energies'' $\epsilon_\nu$,
occupations factors $f_\nu$ at the temperature $T=1/\beta$. Here  $\nu$
includes spin and other relevant quantum numbers.
Quantum systems at high
temperatures behave classically. Then $n(r)$ is given by the
Boltzmann form:
\begin{equation}
n(\vec{r})=n_0(0)\exp(-\beta V_{KS}(\vec{r}))
\end{equation}
where $n_0(0)$ is a reference density, and $V_{KS}$ is a classical Kohn-Sham
potential. This suggests that the
kinetic-energy functional may be side-stepped by (i) the use
of an equivalent ``classical-fluid'' at a temperature $T_{cf}$ for the quantum
system whose actual  physical temperature $T$ may even be zero;  (ii)
use of effective classical  pair-potentials to mimic quantum effects.
In the following we discuss such a ``classical map'', enabling a great
simplification in quantum calculations of interacting electrons.

All quantum observables are mean values over suitable distribution
functions, formed by averaging over most of the variables in the square of
the many-body wavefunction. The most useful averaged quantities are the
one-body and two-body distributions. We had already
demonstrated a simple but accurate  classical map for the interacting
uniform electron fluid (UEF), by presenting explicit calculations of
spin-polarized pair-distribution functions (PDFs) at zero and finite
$T$, for the 3D electron liquid~\cite{prl1,prb}, the 2D-electron
fluid~\cite{prl2,prl3, buluty,totsuji1} and multi-valley
systems~\cite{2valley}. Fermi-liquid parameters of thick-electron layers have
also been determined {\it via } this classical-map technique~\cite{quasi}. 
It has been successfully applied to hot dense hydrogen and related
systems~\cite{cdw-mur,hyd}.

The method employs a {\it classical} Coulomb fluid whose PDFs are determined
{\it via} classical statistical mechanics. The map using the Hyper-Netted-chain
(HNC) method is  named the CHNC. Molecular dynamics (MD) 
simulations may also be used~\cite{cdw-mur}, where it was called the CMMD. The
temperature $T_q$  of the ``equivalent classical Coulomb fluid'' is chosen to
reproduce the correlation energy of the original quantum fluid at $T=0$. Then
the PDFs of the classical fluid at 
$T_{cf}=\surd{(T^2+T_q^2)}$ are excellent approximations to the PDFs of the
quantum fluid at $T$. The so-obtained PDFs are then used in the standard way to
obtain the energies and other properties of the system. Where possible,
the accuracy of the CHNC results have been demonstrated by comparison
with quantum Monte Carlo (QMC) or DFT results. 

Here we apply the CHNC to a typical {\it inhomogeneous}
systems, viz., electrons trapped in 2D parabolic potentials (quantum dots). In
standard calculations, if $N_e$ electrons are in the ``external
potential'', a suitable basis set of $N_b$ functions, with $N_b$  significantly
larger than $N_e$, is selected. A Hartree-Fock (single-determinant) calculation
is followed by a configuration-interaction (CI) expansion in Slater
determinants. The complexity of the problem grows factorially with $N_b$. It is
the electron-electron interactions, which make the
problem prohibitive. In CHNC or CMMD, we treat  many-body
effects classically (i.e., an $\cal{O}$(0) approach), while the non-interacting
Hamiltonian $H_0$ is treated exactly. Here we summarize the salient features of
the classical-map technique: (i) Assignment of a classical-fluid
temperature $T_{cf}$ to the electron system. (ii) Replacement of the
Coulomb-interaction operator $1/\hat{r}$ by a classical ``diffraction
potential'' $v_{c}(r)$= $\{1-\exp(-k_{th}r)\}/r$ which accounts for the thermal
de Broglie length $1/k_{th}$ of the electron at $T_{cf}$. (iii) Ensuring that
the {\it non-interacting} electron PDFs with spin polarization $\zeta$, viz.,
$g^0(r,T,\zeta)$ are correctly  recovered if the Coulomb interaction is
switched off. Thus a ``Pauli exclusion potential'' $P(\vec{r})$ is used to
exactly reproduce the Fermi hole~\cite{lado} in $g^0(r,T,\zeta)$. We apply CHNC
to parabolic quantum dots to show that their interacting inhomogeneous charge
densities and energies can be readily calculated {\it via} CHNC, for arbitrary
$N_e$. 

{\it Quantum dots--} Electrons trapped in parabolic potentials are found in
Fermion traps and in the quantum dots of nanotechnology\cite{ims}. Such quantum dots have been studied extensively by several
methods~\cite{liparini, mario}, including QMC~\cite{popsueva}. Using atomic
units ($|e|=\hbar=m_e=1)$, the Hamiltonian operator $H=H_0+H_{int}$ for
electrons subject to a potential $u(\vec{r})$ is given by
 \begin{equation}
 H_0=\sum_i\left[\frac{\nabla^2_i}{2} +u(\vec{r }_i)\right],\;
 H_{int}=\frac{1}{2}\sum_{i\ne j}\frac{1}{\vec{r}_{ij}}
 \end{equation}
 The classical map converts the Hamiltonian operator to the classical
  Hamiltonian  $H_c$.
 \begin{equation}
 H_c=\sum_i\left[\frac{p^2_i}{2} +u_c(\vec{r }_i)\right]+
 \frac{1}{2}\sum_{i\ne j}\left[v_c(\vec{r}_{ij})+
 P_{s_i,s_j}(\vec{r}_{ij})\right]
 \end{equation}  
 Here $\vec{p}_i$ is the momentum of the $i$-th electron, with spin $s_i$.  The
 Coulomb-interaction {\it operator} $1/\hat{r}$ is replaced by the well-known
 diffraction-corrected classical form $v_c(r)$.   The {\it operator}  defining the
 parabolic confinement in the dot is $u(\vec{r})=$ $\omega_0^2\vec{r}\,^2/2$. It maps
 to the classical function $u_c(\vec{r})$, constructed so that the non-interacting
 density $n^0(r)$ is recovered from $u_c(\vec{r})$ as a classical distribution.
 The potentials $u(\vec{r})$ and $u_c(\vec{r})$ differ
 because the quantum system is sensitive to the boundary conditions imposed on
 the eigenstates of $H_0$

 Thus the essential input to the classical mapping of inhomogeneous systems is
 the {\it non-interacting } density $n^0(\vec{r})$. Here we suppress vector
 notation (unless needed for clarity) and consider circular quantum dots. Given
 $n^0(r)$, we seek the classical potential which generated it. As this
 involves the inversion of an inhomogeneous HNC-type equation, we proceed
 indirectly. If the presence of each electron did not affect any other electron,
 the corresponding classical potential $u_c(r)$ is:
 \begin{equation}
 \label{uclas}
 n^0(r)\equiv n^0(0)\exp\{-\beta u_c(r)\}
 \end{equation}
 This equation determines the product $\beta u_c(r)$, and not separately the
potential $u_c(r)$, or an inverse temperature $\beta$. The reference zero of
all potentials will be referred to the center of the dot.
The classical potential $u_c(x)$ depends on  $N_e$
even though the electrons are
non-interacting. In effect, they have developed indirect steric
interactions {\it via} the confining potential.
 
\begin{figure}[t]
\includegraphics[angle=0, width=3.3in]{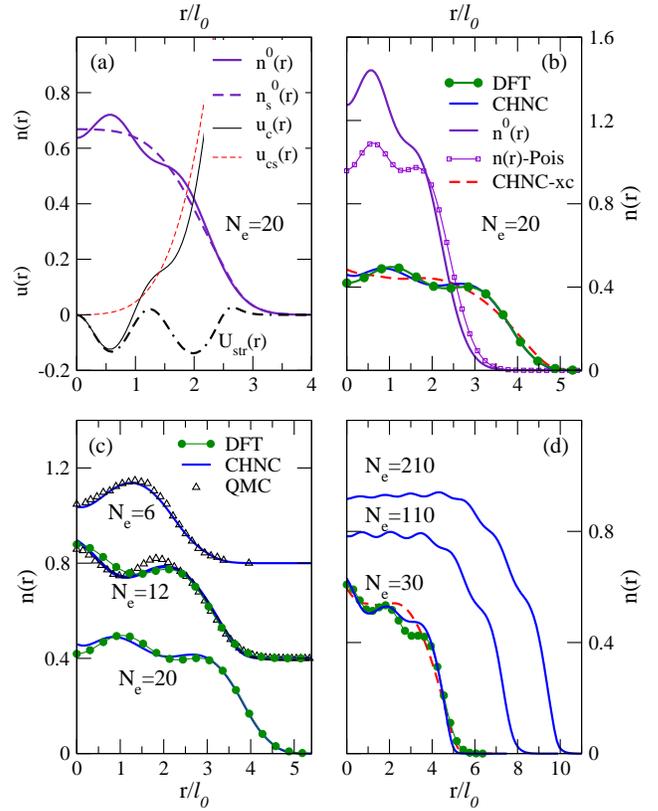}
\caption{(Color online) Electron densities and potentials
for a 2D harmonic-oscillator quantum dot with $N_e$=6,12,20,30,
and 110 and 210 electrons. The
unit of length $\ell_0=1.89$ a.u., or 11 nm, with $\omega_0=0.28$ a.u.,
i.e., 3.3 meV. Panel (a) shows the non-interacting
$n^0(r)$, and also a smoothed charge density $n^0_s(r)$. The corresponding
classical potentials $u_c(r)$ and $u_{cs}(r)$ are also shown. The
difference potential, $U_{str}(r)$, defines a ``steric -packing'' 
potential which is used to construct a Bridge correction $B(r)$ to the HNC
equations. Panel (b) shows the evolution of $n^0(r)$ to the interacting
$n(r)$ with the inclusion of Poisson, exchange and
correlation (xc) terms, to give CHNC-xc, which does not
contain $B(r)$, and CHNC inclusive of $B(r)$.
In (c) the final CHNC results
(inclusive of bridge terms) for $N_e$=6,12, and 20 are given.
In (d) results for $N_e$=30, and 110 and 220 are
shown. The QMC data points were extracted from Ref.~\cite{mario}.  
}
\label{qdotfig1}
\end{figure}

The confining potential
defines a length scale $\ell_0=\surd{(\hbar/(m\omega_0))}$.
We use the effective mass $m^*=0.067$ and the
dielectric constant $\kappa=12.4$ typical of GaAs. These define
 effective atomic units (a.u.) with a Hartree energy of
 $me^4/(\kappa\hbar)^2\simeq 11$ meV, and a Bohr radius
 $a_B=\hbar^2\kappa/(me^2)\simeq 9.79$ nm.
 In Fig.~\ref{qdotfig1}(a) we show the $n^0(r)$ and $\beta u_c(r)$ for
 a 20-electron circular quantum dot, $N_e=20$,
 with $\omega_0=3.33$ meV~\cite{note1}.
 The non-interacting density $N^0(r)$ shows shell-filling effects. In
Fig.~\ref{qdotfig1}(a) we show a smoothed charge density $n^0_s(r)$ as well. The
density difference  $n^0(r)-n^0_s(r)$ integrates to zero, and may be thought of as
the result of a steric interaction $U_{str}(r)$ arising from the packing of
classical particles into the parabolic trap.  In the quantum system it arises from
the  boundary conditions on $H_0$. As many shells are filled with increasing $N_e$,
and as Coulomb interactions come on, this effect becomes unimportant, as seen in the
interacting-$n(r)$ for the dot with $N_e$=210, in Fig.~\ref{qdotfig1}(d). 
 
 The electron-distribution $n_\alpha(r)$, spin $\alpha$, in the dot
 defines a dot-electron PDF by, e.g., $n_\alpha(r)=n_\alpha(0)g_{d\alpha}(r)$.
 This in turn depends on  the electron-electron PDFs $g_{ee}$,
  i.e., explicitely,
 $g_{\alpha,\alpha'}(\vec{r},\vec{r}\,')$. Evaluating this coupled set is
 complex even within classical mechanics. Unlike in {\it uniform} systems, 
 these PDFs dependent on $\vec{r},\vec{r}\,'$. While a classical-map-MD
 approach is the best option, we show that simple approximations to the CHNC
 integral equations lead to surprisingly accurate results.

 Our simplified approach is based on constructing the density
 $n_{d\alpha}(r)=n_\alpha(0)g_{d\alpha}(r)$ while the e-e PDFs  
 $g_{\alpha,\alpha'}(\vec{r},\vec{r}\,')$ are evaluated from an average-density
 approximation. This saves us from solving a set of coupled HNC equations. The
 approximations proposed are: (i) replacing the e-e PDFs
 $g_{\alpha,\alpha^\prime}(\vec{r},\vec{r}\,')$ by the PDFs of a uniform slab
 (USB) of average density, $\bar{n}$, (ii) determining the USB density, viz.,
 $\bar{n}$ from $<n(r)n(r)>$/$<n(r)>$, as in Refs.~\cite{quasi,ggsavin}, (iii)
 using $\bar{n}$ and the equations of Ref.~\cite{prl2} to determine $T_q$ and
 other UEF parameters needed for the CHNC, (iv) calculating the charge density
 $n_{d\alpha}(r)$ in the quantum  dot {\it via} a simplified HNC-like classical
 integral equation inclusive of a bridge term $B_{de}(r)$. For brevity of
 presentation, we use a spin-unpolarized system ($\zeta=0$), and write
 $g_{de}(r)$ for the PDF defining the charge density in the dot. The basic
 CHNC equation for the electron density in the inhomogeneous system is:
 \begin{equation}
 \label{doteqn}
 n(r)=n(0)\exp\left[-\beta u_c(r)+ V_{m-b}\right]
 \end{equation} 
 This is a Boltzmann distribution for the pair-potential $u_c(r)$
 inclusive of its many-body correction $V_{m-b}$. This includes mean-field and
 correlation effects. The classical mean-field term is just a Poisson potential. We
 also need the correlation potentials beyond mean-field from classical mechanics.
 Hence we rewrite $V_{m-b}$ in terms of the Nodal function $N(r)$ and the bridge
 function $B(r)$ of HNC theory, remembering that ``exchange'' already appears in the
 classical map as an effective pair-potential (``Pauli potential'') between
 like-spins. The ``bridge'' diagrams brings in irreducible (three-body and higher)
 packing effects beyond simple HNC. 
 \begin{eqnarray}
 \label{dotset}
 V_{m-b}&=&N(r)+B_{de}(r)\\
 N(r)&=&\beta\left[V_p(r) + V_x(r) +V_c(r)\right]\\
 V_p(r)&=&\int n(r')d\vec{r'}/|\vec{r}-\vec{r'}|\\
 V_x(r)&=&\int n(r')d\vec{r'}P(\vec{r},\vec{r'})\\
 V_c(r)&=&\int n(r')d\vec{r'}\left[\log\{g_{ee}(\vec{r},\vec{r'})\}
 -h_{ee}(\vec{r},\vec{r'})\right]
 \end{eqnarray}
 The nodal term $N(r)$ has been decomposed into a Poisson potential $V_p(r)$,
 an ``exchange potential'' $V_x(r)$ arising from the Pauli exclusion potential
 $P(\vec{r},\vec{r}\,')$, and a classical correlation potential $V_c(r)$. This is
 a standard analysis based on the Ornstein-Zernike equation, and was already
 discussed in Ref.~\cite{hyd1}. The expression for $V_c(r)$ is Eq.~(3.4) given
 there, and $h_{ee}(\vec{r},\vec{r}\,')$ is the total correlation function. $V_p$
 has been written with the Coulomb potential $1/r$ rather than with the
 diffraction correction (DC), since we found that the numerical effects of DC
 are negligible for the $g_{de}(r)$ of the quantum dots studied here (n.b. the
 DC is needed in the CHNC evaluation of the $g_{ee}(r)$ at the uniform-slab
 density $\bar{n}$).
 
We approximate $g_{ee}(\vec{r},\vec{r}\,')$ by $g_{ee}(|\vec{r}-\vec{r}\,'|)$ of
the uniform slab at density $\bar{n}$. The $\bar{n}$ in our quantum dots
are found to have Wigner-Seitz radii $r_s\sim 1$ and hence the Coulomb
correlation potential $V_c(r)$ is small, while the Pauli exclusion effect is
large. Here $P(\vec{r},\vec{r}\,')$ is replaced by $P(|\vec{r}-\vec{r}\,'|)$ at the
uniform density $\bar{n}$. This is a universal function of $x=r/r_s$. It can be
fitted to $\{-a2\log(x)+b\}/(1+cx)$, or to the simpler form $a_1/(1+a_2x)$,
with $a_1$=10.1186, $a_2=3.69352$, for 2D systems at $T=0$. Replacing 
$g_{ee}(\vec{r},\vec{r}\,')$ by its uniform-density value leads to the
question of the appropriate form for $n(\vec{r}\,')$ which now has to play the
role of a $n(\vec{r},\vec{r}\,')$. The usual simple choices (e.g.,
$n(\vec{r}\,')\to (n(\vec{r})+ n(\vec{r}\,'))/2$, or even  $n(\vec{r})$ give
similar  results, but the  replacement $n(\vec{r}\,')\to \bar{n}$,
i.e., completely by a uniform slab, is too drastic.
 
   The results obtained from the self-consistent solution of Eq.~\ref{doteqn} are
 shown in Fig.~\ref{qdotfig1}. Panel (a) shows the input potential $\beta u_c(r)$
 based on the non-interacting density $n^0(r)$. This $n^0(r)$ gets modified by the 
 interactions. The average density $\bar{n}$, the corresponding inverse temperature
 $\beta$, potentials $V_p, V_{xc}$ etc., were calculated from $n(r)$ in each
 iteration, with the total number of electrons fixed to $N_e$. In panel (b) the
 self-consistent $n(r)$ obtained as we successively add Poisson (curve with boxes),
 exchange-correlation (dashed red curve), and Bridge corrections (solid blue curve)
 is found to converge to the benchmark results from QMC and DFT (solid green
 circles). Panels (c), (d) shows the full CHNC results and QMC and/or DFT results
 for $N_e$=6,12,20, 30. These have been done using the quantum temperature $T_q$
 defined by Eq. 5 of Ref.\cite{prl2}, which assignes a $_q$ to a given $r_s$. For high
 electron densities when $r_s <1$, the Buluty-Tanatar(BT) map\cite{buluty} seems
 to be more accurate. Calculations for $N_e$=110, 210 etc, using a BT-type map
 are given in panel~(d), but no published QMC or DFT results are available..  
 
 An oscillatory structure in $g_{ee}(r)$ occurs even in uniform fluids, when
interactions are important, and is well understood. The $g_{ee}(r)$  of the uniform
fluid at strong coupling could be accurately recovered on including bridge
contributions $B_{ee}(r)$ to the HNC, as shown in Ref.~\cite{prl2, rosen}. There
the particle-packing theory of the {\it hard-sphere fluid} could be used, since the
PDF is not too sensitive to the details of the bridge interaction.  In the
classical map of the quantum dot, packing effects are dominated by the steric
crowding effect of the confining potential. This
steric potential $U_{str}$ is already available to us in the charge distribution
$n^0(r)$ of $H_0$. We assume that the difference between the smoothed distribution $n^0_s(r)$,
and $n^0(r)$,
Fig~\ref{qdotfig1}(a), corresponds to the effect of $U_{str}(r)$. Then, 
converting a charge distribution into a potential {\it via }the
classical map, we write
 \begin{eqnarray}
 \beta U_{str}(r)&=&-\log\{n^0(r)/n^0(0)\}
  -\log\{n^0_s(r)/n_s^0(0)\}\nonumber\\
 B_{de}(r)&=&\gamma(\beta/\beta_0)\xi^2\gamma V_s(r/\xi),\;\;\xi=r_s/r_s^0
 \end{eqnarray}
 The equation  for $B_{de}(r)$ reflects the rescaling of the uniform-slab density
 $\bar{n}^0$ to $\bar{n}$ due to interactions, changing the scales of the
 parameters $\beta^0,\, r_s^0$, etc., to $\beta, \, r_s$ etc., of the final
 self-consistent density, thus rescaling the steric potential of the
 non-interacting system.  The numerical factor $\gamma$ is set to 1.5. This
 simple model avoids the complex microscopic calculation of a bridge correction, 
 and is seen to be justified {\em a postiori}. It does not
 appeal to any parameterizations outside the problem. In fig.~\ref{qdotfig1}(c) we
 show comparisons of the CHNC $n(r)$ for $N_e=$6,12,20,  with QMC results. DFT
 results for $N_e=20$ are shown in fig.~\ref{qdotfig1}(b). Panel (d) shows the CHNC
 results for $N_e$=30, 110 and 220 electrons. In the last two cases we do not have
 microscopic calculations for comparison with CHNC.
  The calculation of the interacting density $n(r)$ for
 arbitrary $N_e$, at finite temperatures,  finite values
 of $\zeta$ or finite magnetic fields pose no additional difficulty in CHNC.
 
\begin{table}
\caption{The Exchange-Correlation and kinetic energies evaluated from
  the densities (Fig.~1), i.e.,  CHNC $n(r)$, and from the DFT $n(r)$ of
   Ref.~\cite{mario}. $N_e$ is the number of electrons. The energy unit is $\omega_0$.
   }
\begin{ruledtabular}
\begin{tabular}{ccccc}
$N_e$ & $E_{kin}$CHNC &$E_{kin}$DFT & $-E_{xc}$CHNC & $-E_{xc}$DFT\\
\hline \\
6   &2.317            &2.415	    &7.491    &7.638 \\ 
12  &5.981            &5.897	    &16.97    &16.80 \\  
20  &11.55            &11.47	    &30.20    &30.07 \\  
30  &21.30            &19.51	    &49.90    &48.78 \\  
110 &111.2            & --	    &213.1   & --   \\
210 &257.4            & --	    &445.7   & --   \\  
\end{tabular}
\end{ruledtabular}
\label{energytable}
\end{table}
The charge distributions of CHNC can now be used for the total energy $E$, 
which involves the confinement energy $E_c$, the Possion energy $E_{poi}$,
$E_{xc}$, and $E_{kin}$. The exchange-correlation and kinetic contributions
are the quantum mechanically sensitive, ``difficult'' terms.
The simplest approach is found to be adequate for 2D quantum dots. That is,
we use the LDA (local-density approximation), with the known 2D
exchange-correlation energy functionals \cite{atta}. The success of the LDA for
the 2D kinetic energy has been noted by van Zyl et al.\cite{zyl}, and also by
Koivisito et al\cite{PerrotH}.
A comparison of our xc-energies with those from DFT are given in Table~I. 
 
   In conclusion,  we have presented classical-map calculations for a 2D
 inhomogeneous system of {\it interacting} electrons, viz., quantum dots, which are
 in good agreement with microscopic calculations where available. This method
 requires no basis sets, no evaluation of matrix elements etc. It is an order-zero,
 viz., $\cal{O}$(0) approach independent of the number of electrons. Similar
 applications to atomic systems, (``naturally occurring quantum dots''), are
 clearly feasible. The uniform-slab  approximation, and the need to model the
 bridge-diagram corrections to the CHNC equations may be avoided by resorting to a
 classical molecular-dynamics implementation of the method\cite{miyake}. 
 The author thanks Mario
 Gattobigio for providing the DFT $n(r)$ and energies of Ref~\cite{mario}.

\end{document}